\documentclass[cits]{PoS}
\usepackage{multirow,xspace}

\newcommand{\be}{\begin{equation}}
\newcommand{\ee}{\end{equation}}
\newcommand{\q}{\quad}

\renewcommand{\slash}[1]{#1\!\!\!\!\!/}
\newcommand{\MET}{$\slash E_T$\,}
\newcommand{\Ho}{{\tilde H}}

\newcommand{\st}{{\tilde t}}

\newcommand{\snu}{{\tilde\nu}}

\newcommand{\stau}{{\tilde\tau}}

\title{LHC searches examined via the RPV MSSM}

\ShortTitle{LHC searches examined via the RPV MSSM}

\author{Jared A. EVANS$^a$\, and \speaker{Yevgeny KATS}$^{ab}$\\
        \llap{$^a$}New High Energy Theory Center, Rutgers University, Piscataway, NJ 08854, USA\\
        \llap{$^b$}Dept.\ of Particle Physics and Astrophysics, Weizmann Institute of Science, Rehovot 76100, Israel\\
        E-mail: \email{jaevans@physics.rutgers.edu}, \email{yevgeny.kats@weizmann.ac.il}}

\abstract{We confront the ATLAS and CMS new physics searches with simplified models from $R$-parity violating (RPV) supersymmetry that contain naturally light stops and higgsinos. We emphasize the important role played by searches for exotic heavy quarks and leptoquarks and discuss how these searches could be modified to better address RPV scenarios. We also argue that it would be useful to design searches for new physics based on the $t\bar t$ cross section measurements, to enhance searches for taus + jets + \MET with regions with $b$ tagging and reduced \MET requirements, and to expand the search program for all-jet final states, especially in the presence of multiple $b$-jets.}

\FullConference{The European Physical Society Conference on High Energy Physics \\
                 18-24 July, 2013\\
                 Stockholm, Sweden}

\begin{document}

\section{Introduction}

If the electroweak scale is stabilized by supersymmetry (SUSY), then the higgsino masses should be around the Higgs mass, and top squarks (stops) should not be much heavier. By now, many SUSY models have been addressed either directly by ATLAS or CMS, or via reinterpretations of their searches by theorists. These studies have excluded a large range of scenarios, including many conservative ones in which stops are the only abundantly produced superpartners. However, some scenarios were identified where existing LHC searches were insensitive. Most relevant to the present note is our earlier work~\cite{our-stops}, which classified the signatures of the various possible decays of stops in $R$-parity violating (RPV) models~\cite{RPV-review}, and analyzed them in the context of LHC searches. The multitude of possible RPV couplings (including their flavor structure) lead to a diverse set of scenarios, many of which were found to elude the LHC searches available at the time of~\cite{our-stops}. In this note, we will focus on the developments in LHC searches that have occurred since then and point out several additional promising search strategies that ATLAS and CMS could implement.

The RPV scenarios we will examine here are potentially challenging for LHC searches both because the missing energy is low (arising only from top or tau decays, if it is present at all), and because in some cases the only sources of electrons and muons are taus, or there are no leptons at all. On the other hand, there are also opportunities, that are not fully being used, to probe these scenarios. First, the jet multiplicity is typically very high, due to both cascade decays and RPV couplings involving quark superfields. Second, multiple $b$-jets are typically present, due to both third-generation squarks and specific RPV couplings. Finally, in most cases, some electrons and muons \emph{are} available, even if only from top or tau decays.  We will examine how sensitive existing searches are to such scenarios and what extensions to the search strategies these scenarios motivate.

\section{Benchmark models}

\begin{figure}[t]
\begin{center}
\begin{picture}(0,0)(0,0)
\large
\put(-5,50){$\st$}
\put(150,100){$b$}
\put(150,65){$f_2$}
\put(150,35){$f_3$}
\put(150,0){$f_1$}
\put(95,50){$\tilde f$}
\put(50,25){$\Ho$}
\normalsize
\end{picture}
\includegraphics[scale=1.3]{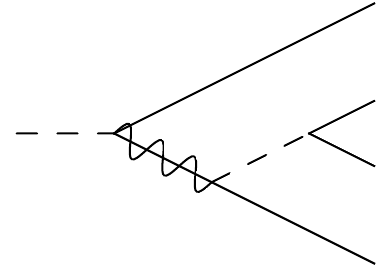}
\caption{Stop decay processes. The higgsinos $\Ho$ are assumed to be on-shell, while the sfermion ($\tilde f = \st_R$, $\st_L$, $\stau_L$, or $\snu_\tau$) is assumed to be off-shell. The symbols $f_1$, $f_2$, $f_3$ denote Standard Model fermions.}
\label{fig:diagram}
\end{center}
\end{figure}
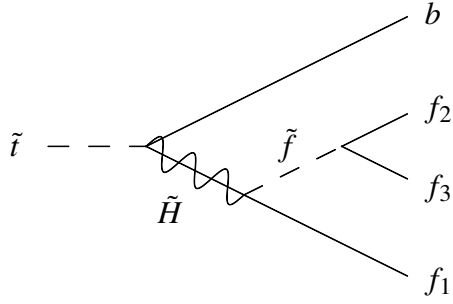

The examples we will analyze involve the lightest stop ($\st$) being pair-produced before decaying to a higgsino ($\Ho$), namely
\be
pp \to \st\st^\ast \,,\q \st \to b \Ho^+ \,.
\ee
Here we have assumed $m_\st - m_\Ho < m_t$ so that the decays $\st \to t\Ho^0_{1,2}$ are absent. The chargino $\Ho^+$ can decay either through the lightest neutralino $\Ho^0_1$
\be
\Ho^+ \to W^{+\ast} \Ho^0_1 \,,\q W^{+\ast} \to \mbox{soft particles} \,,\q \Ho^0_1 \to \mbox{RPV}
\label{eq:transition}
\ee
(where we assume $m_{\Ho^+} - m_{\Ho^0_1} \lesssim$~few~GeV, so that the decay products of the off-shell $W^+$ are essentially unobservable) or directly
\be
\Ho^+ \to \mbox{RPV} \,.
\label{eq:no-transition}
\ee
The RPV decays proceed through a sfermion (slepton or squark) $\tilde f$ and a certain $UDD$ or $LQD$ coupling.\footnote{$LLE$ couplings are less interesting to us because they are easily excluded by multilepton searches~\cite{our-stops}.} The process is illustrated schematically in Fig.~\ref{fig:diagram}. Table~\ref{tab:scenarios} describes the sfermion and coupling choices, the assumption of whether (\ref{eq:transition}) or (\ref{eq:no-transition}) occurs, and the resulting stop decay final states for each scenario we will discuss.  Note that in the case of (\ref{eq:transition}), leptons from stop and antistop decays can have the same sign. In all cases, we will take
\be
m_\Ho = m_\st - 100\mbox{ GeV} \,,\q
m_{\tilde f} \gg m_\Ho.
\ee
We will also assume the mass splitting between $\Ho^+$ and $\Ho^0_1$, the size of the RPV coupling, and the mass of $\tilde f$ to be such that all the decays are prompt.

\begin{table}[t]
\begin{center}\small{
\begin{tabular}{|c|c|c|c|c|c|}\hline
&  \multicolumn{3}{|c|}{Scenario}                       & $\Ho^+ \to \Ho^0_1$ & Final state \\\cline{2-4}
&  \multicolumn{2}{|c|}{Coupling} & Mediator $\tilde f$ & transition          & (for each stop) \\\hline
A & \multirow{2}{*}{$UDD$}
    & 312 & $\st_R$        & YES & $tbqq$ \\\cline{3-6}
B & & 323 & $\st_R$        & NO  & $bbbq$ \\\hline
C & \multirow{3}{*}{$LQD$}
    & 323 & $\stau_L$      & YES & $\tau bbq$ \\\cline{3-6}
D & & 321 & $\snu_\tau$    & NO  & $\tau bqq$ \\\cline{3-6}
E & & 232 & $\st_L$        & NO  & $\mu bbq$  \\\hline
\end{tabular}}\end{center}
\caption{Benchmark models (referred to later by their labels A--E).}
\label{tab:scenarios}
\end{table}

\section{Simulation of Searches}
\label{sec:searches}

Our simulation framework includes a comprehensive set of ATLAS and CMS searches from the 7 and 8~TeV LHC runs. Among the more recent searches added (relative to~\cite{our-stops}), we would like to mention in particular the exotic heavy quark searches from ATLAS~\cite{ATLAS-LSST-3b} and CMS~\cite{CMS-B2G},\footnote{The single-lepton channel of the CMS search uses a boosted decision tree and is therefore not useful for reinterpretation studies like ours. The opposite and same-sign dilepton channels are included in our framework.} the CMS search for second-generation leptoquarks~\cite{CMS-LQ2}, and the ATLAS search for 6-7 high-$p_T$ jets~\cite{ATLAS-6-7}.  In addition to discussing existing searches and ways in which they could be strengthened in the context of our scenarios, we will examine the expected sensitivity of a search, proposed in~\cite{LSST}, for high-$S_T$ events with a lepton, high jet multiplicity and a $b$ tag (without a significant \MET requirement). We have implemented this search as in Sec.~3.3 of~\cite{our-gluinos}, assuming 20~fb$^{-1}$ at 8~TeV.

For generating events with RPV decays, we used \textsc{MadGraph~5}~\cite{MG5} with the RPVMSSM model~\cite{RPVMSSM_UFO}, and then \textsc{Pythia~8}~\cite{Pythia8}. For the technical details of how we simulate searches, see the appendix of~\cite{our-gluinos}. In all cases, we have applied an efficiency threshold of $\sim 10^{-3}$ to reduce sensitivity to the far tails of the signal distributions.

\section{Results}

\subsection{Tops/taus + many jets (with many $b$ tags)}

\begin{figure}[t]
\begin{center}
\includegraphics[scale=0.57]{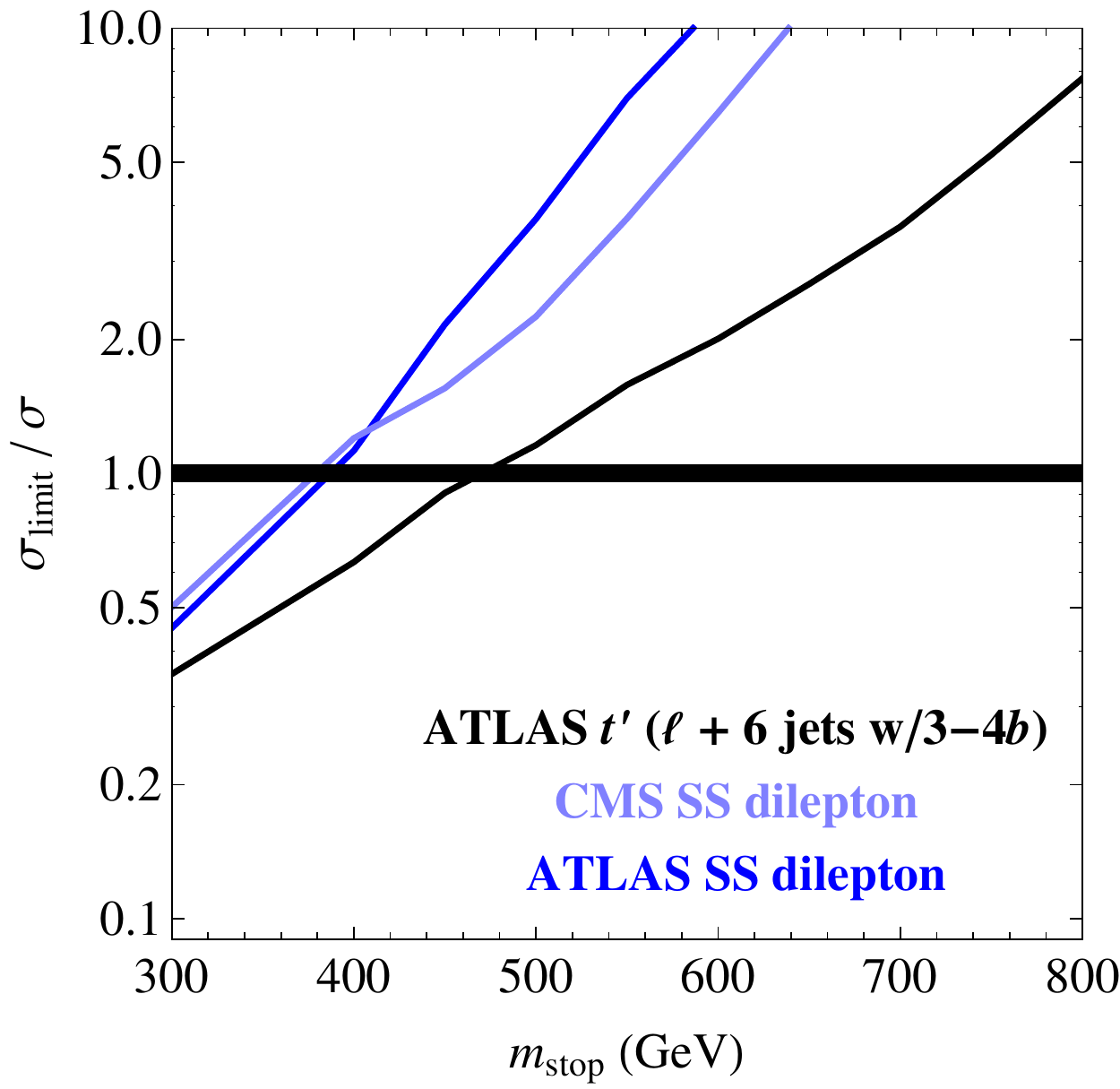}
\includegraphics[scale=0.57]{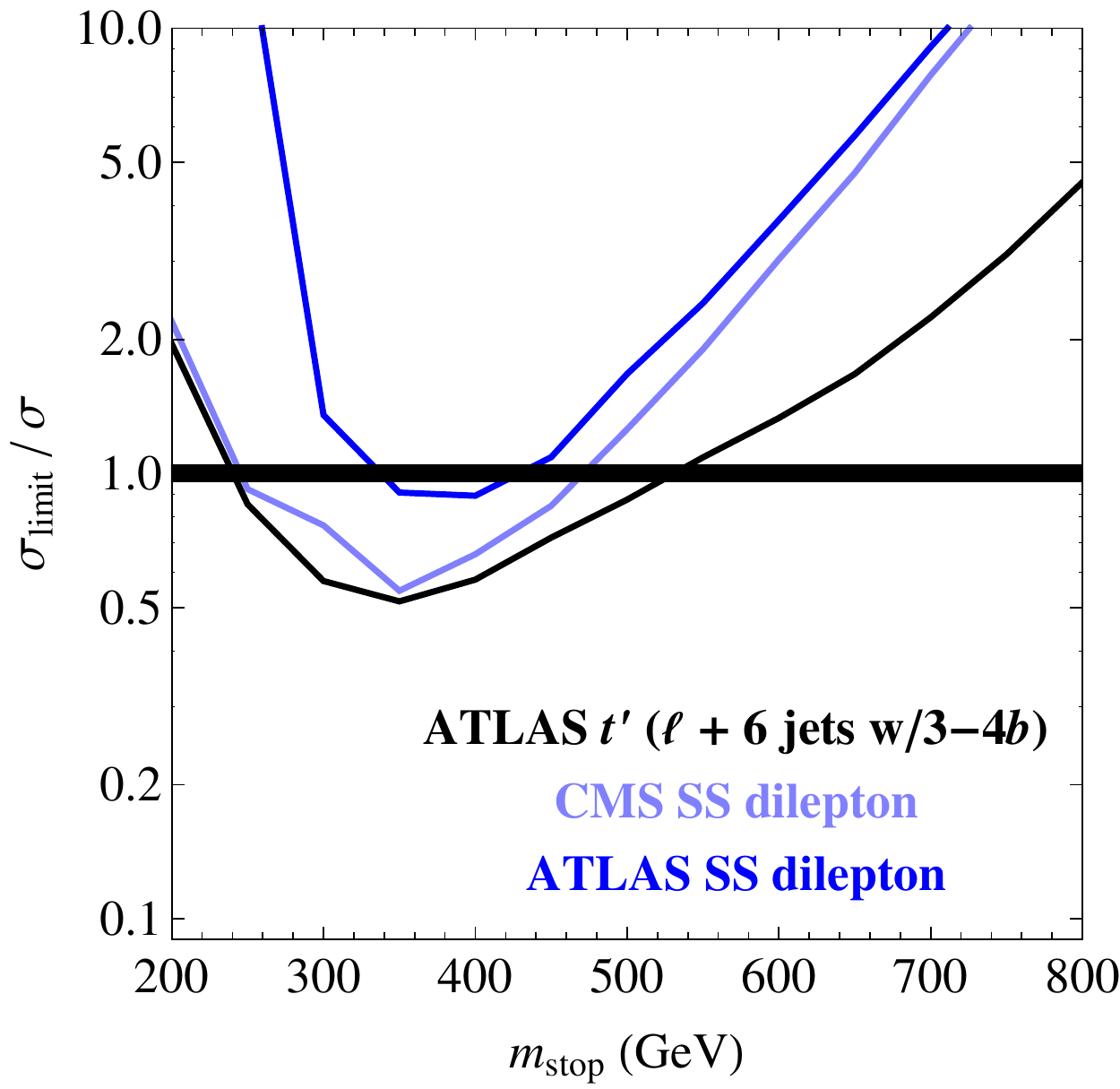}
\caption{Limits on model A, where $\st \to tbqq$ (left) and model C, where $\st \to \tau bbq$ (right), from~\cite{ATLAS-LSST-3b,CMS-SSDIL,ATLAS-SSDIL}.}
\label{fig:tops-taus-manyb}
\end{center}
\end{figure}

In Fig.~\ref{fig:tops-taus-manyb} (left), we present limits on a scenario in which the stop decays often contain leptons (from top decays) and multiple jets. Note that even though SS dileptons are available, the strongest limits come from an ATLAS search~\cite{ATLAS-LSST-3b}\footnote{For our reinterpretation of this search, we defined the search regions as $H_T > 800$, 1000, 1200, 1400, 1600, 1800~GeV for bins with 3 and $\geq 4$ $b$-tags (on top of ATLAS's selection).} that requires a lepton, large ($\geq 6$) jet multiplicity and multiple $b$ tags (but not much $\slash E_T$). It is likely that the reach of this search could be improved by using bins with more than 6 jets, since the signal contains 10 hard partons.

In Fig.~\ref{fig:tops-taus-manyb} (right), we show that the same search~\cite{ATLAS-LSST-3b} excludes a certain range of masses even when the leptons are only coming from tau decays. It would be useful to design a search optimized for lower values of $H_T$, to cover the gap at low stop masses.

\begin{figure}[t]
\begin{center}
\includegraphics[scale=0.62]{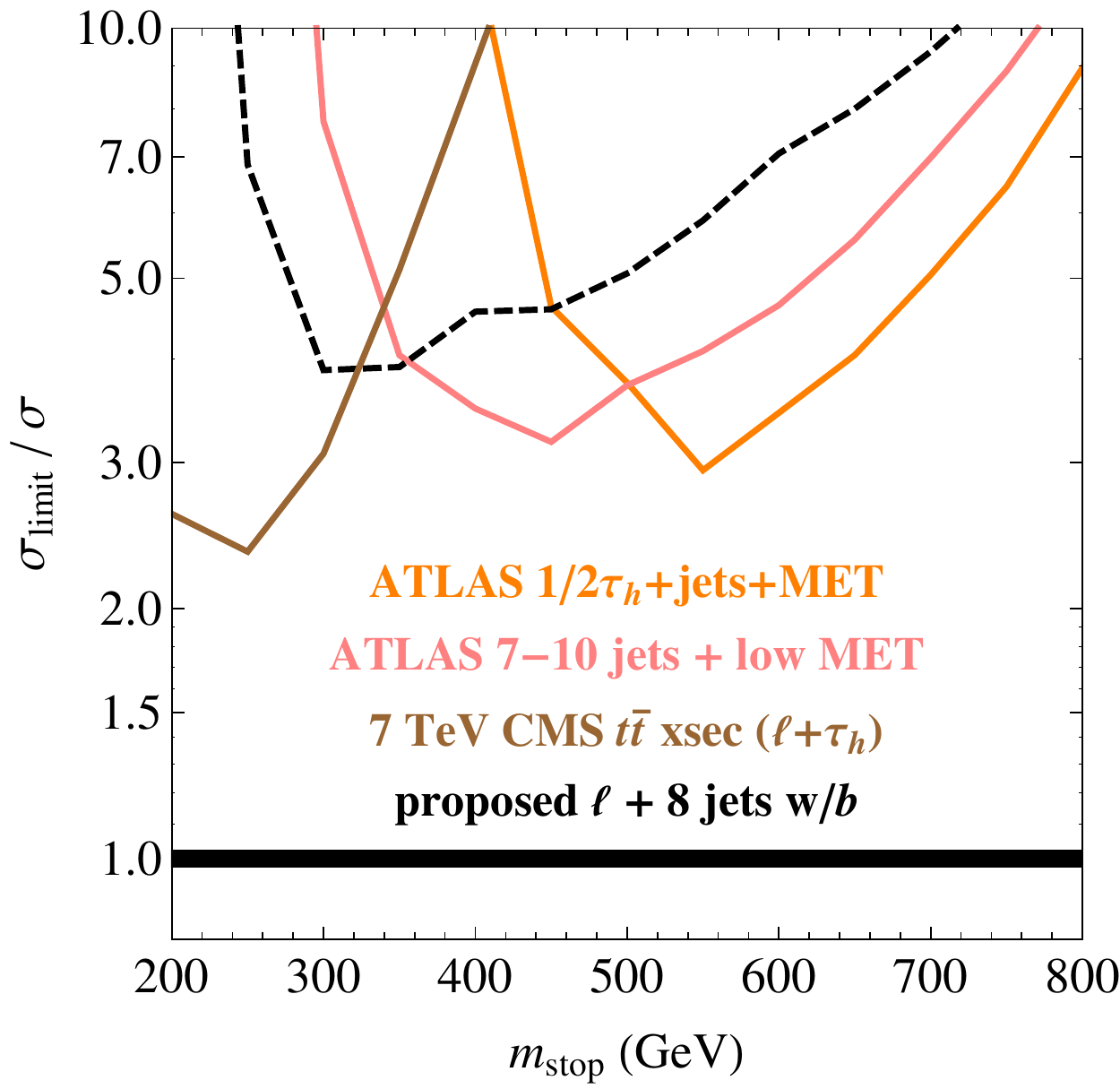}
\caption{Limits on model D, where $\st \to \tau bqq$, from~\cite{ATLAS-SUSY-taus,ATLAS-7-10,CMS-ttbar-tau} and an expected limit (dashed) from a lepton + many jets search proposed in~\cite{LSST,our-gluinos}.}
\label{fig:taus-fewb}
\end{center}
\end{figure}

\subsection{Taus + many jets (with few $b$ tags)}

In Fig.~\ref{fig:taus-fewb}, we present results for another tau-rich scenario, but this time with only one $b$-quark per stop decay. The ATLAS search~\cite{ATLAS-LSST-3b}, which requires 3 or 4 $b$ tags, is no longer useful. However, the proposed search for a lepton + many jets with a \emph{single} $b$ tag~\cite{LSST,our-gluinos} (which we mentioned in section~\ref{sec:searches}) is not very far from being sensitive, especially at low masses. Apart from optimizing the event selection cuts of that search (which we have not attempted), one could try designing an analogous search for a lepton + hadronic tau + many jets.

We also see in that plot that the ATLAS searches for hadronic taus + jets + \MET~\cite{ATLAS-SUSY-taus} and large jet multiplicity + low \MET~\cite{ATLAS-7-10} are not far from being sensitive at intermediate masses. One could likely obtain better sensitivity by using $b$ tagging and lowering the \MET cuts in~\cite{ATLAS-SUSY-taus} (since the \MET in our scenario comes only from tau decays), or alternatively by adding a hadronic tau requirement in~\cite{ATLAS-7-10} (since all the events contain two taus).

Another approach to probing this scenario would be to modify the CMS search for third-generation leptoquarks and RPV stops decaying to $\tau b$ (available in~\cite{CMS-LQ3} with 7~TeV data) in a way that would account for the different kinematics of the $\tau$s and $b$s and utilize the additional jets.

Finally, note that at very low masses the limit from the $t\bar t$ cross section measurement (in the $\ell+\tau_h$ channel)~\cite{CMS-ttbar-tau}, even just with $\sim 2$~fb$^{-1}$ at 7~TeV, is stronger than the limits from any of the new physics searches that we examined. Recasting the $t\bar t$ cross section measurement, perhaps with the requirement of a larger-than-usual jet multiplicity, as a new physics search, would be useful.

\subsection{Opposite-sign dileptons + many jets}

\begin{figure}[t]
\begin{center}
\includegraphics[scale=0.57]{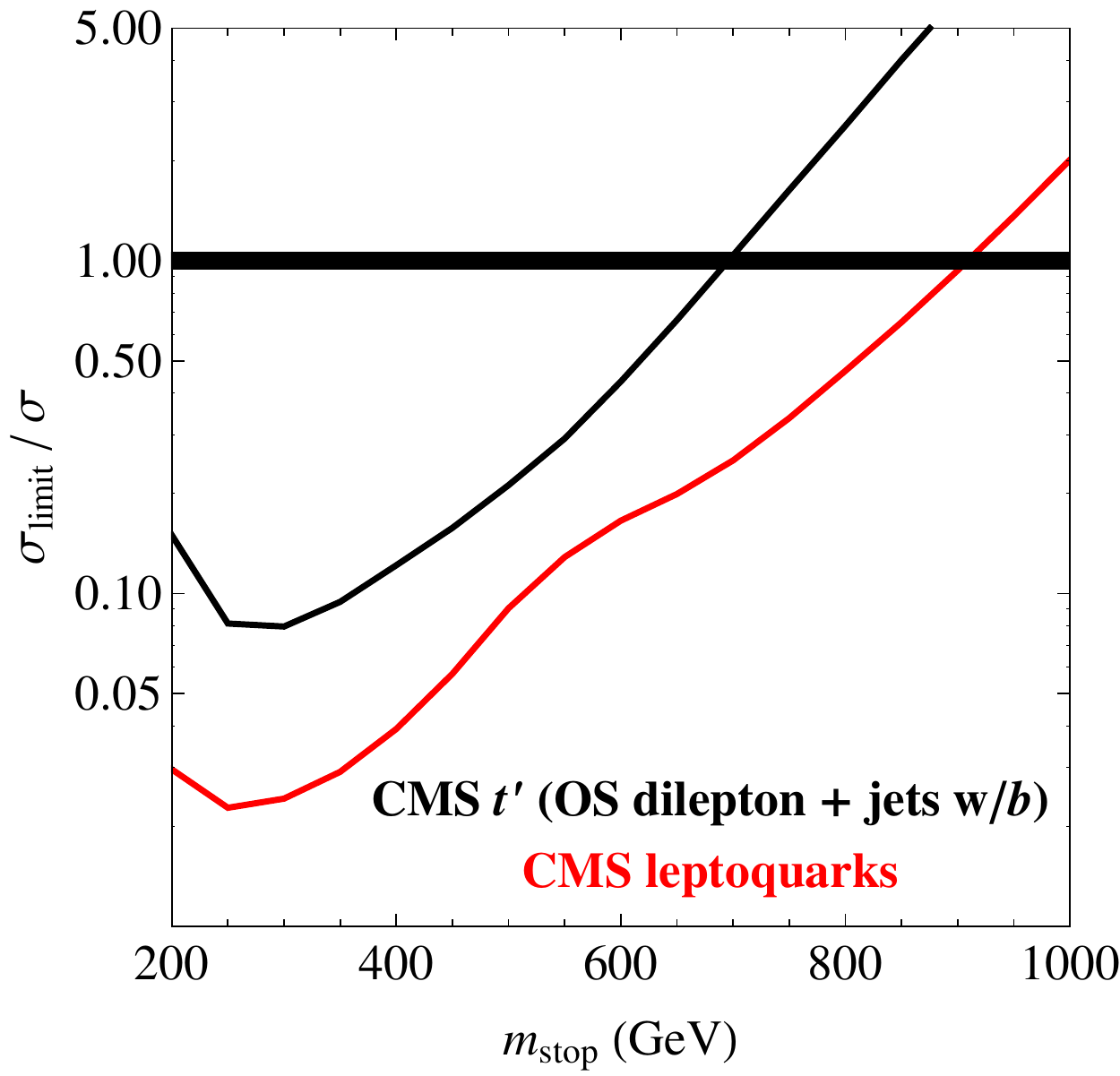}
\includegraphics[scale=0.57]{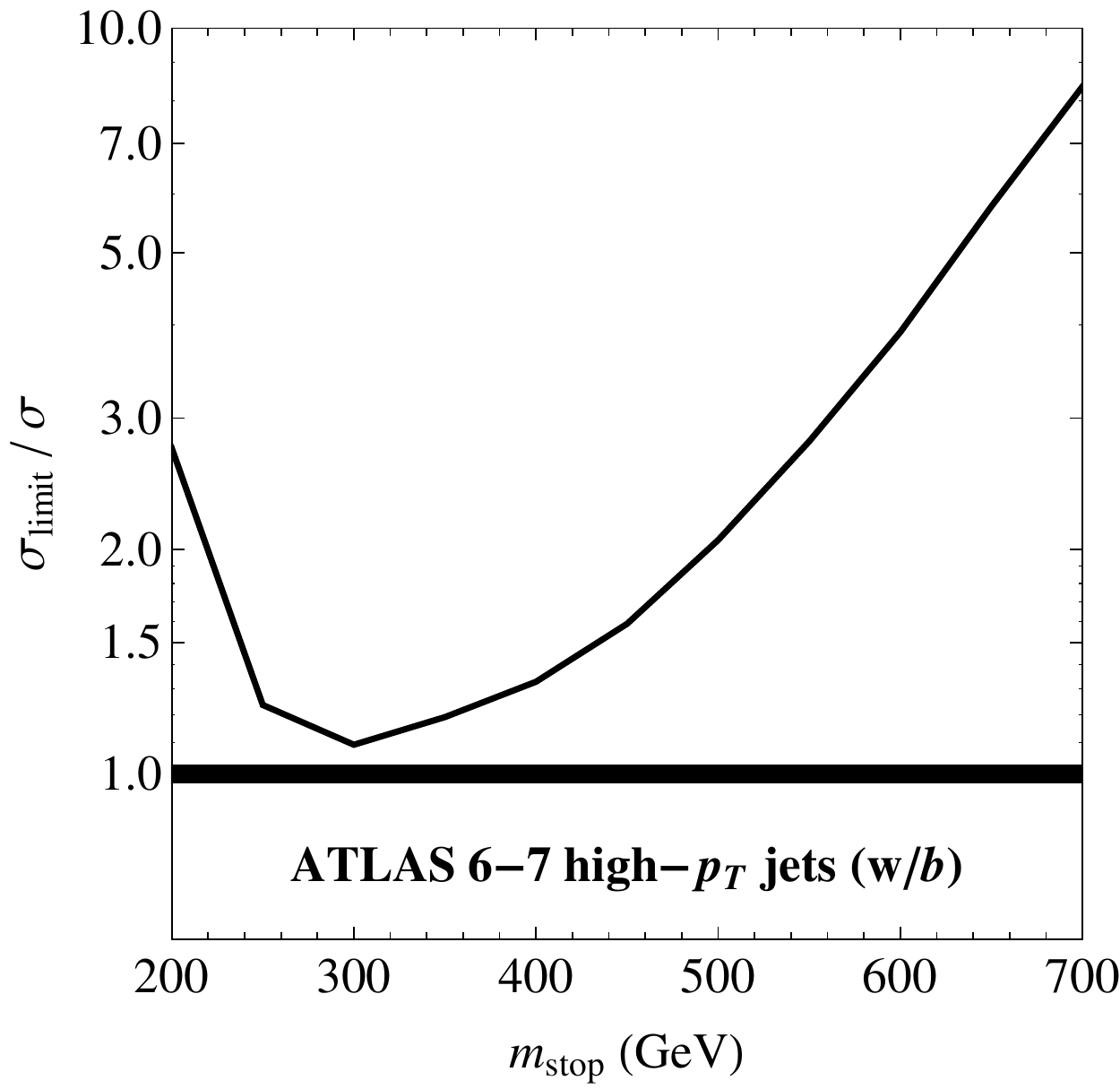}
\caption{Left: limits on model E, where $\st \to \mu bbq$, from~\cite{CMS-B2G,CMS-LQ2}. Right: limits on model B, where $\st \to bbbq$, from~\cite{ATLAS-6-7}.}
\label{fig:LQ-bbb}
\end{center}
\end{figure}

In Fig.~\ref{fig:LQ-bbb} (left), we show that the CMS search for second-generation leptoquarks~\cite{CMS-LQ2} delivers very strong limits on scenarios with pairs of opposite-sign leptons, even though the search's motivation is a two-body rather than a four-body decay. Here we have taken advantage of the fact that CMS uses a separate set of cuts for each leptoquark mass hypothesis. In our reinterpretation, we have used all of these as potential search regions, regardless of the stop mass, in order to allow for the different kinematics. The limits could likely be strengthened by making use of the large jet multiplicity and/or the presence of $b$ jets in the signal. Limits from the relevant opposite-sign dilepton region of the CMS search for exotic heavy quarks~\cite{CMS-B2G}, which does not utilize the high $p_T$ of the leptons and has just a single $S_T$ bin, are somewhat weaker.

\subsection{Multiple $b$ jets}

In Fig.~\ref{fig:LQ-bbb} (right), we consider a scenario with a jets-only final state. While all-jet scenarios with stop production cross sections are generally difficult, in this particular scenario 6 of the 8 jets are $b$-jets. Clearly, the background for such events is small, but there is no search that utilizes this fact. The ATLAS search for 6-7 high-$p_T$ jets~\cite{ATLAS-6-7}, which almost sets a limit at low masses, uses bins requiring 0, 1 or 2 $b$-tags. Adding bins with larger numbers of $b$-tags would be beneficial. More detailed kinematic properties of the signal, such as resonant structures, could also be exploited.

\section{Conclusions}

By examining a set of new physics benchmark models, we have identified certain well-defined final states that are not optimally covered by existing LHC searches. We have proposed strategies, which are modifications on existing searches for exotic heavy quarks~\cite{ATLAS-LSST-3b,CMS-B2G}, leptoquarks~\cite{CMS-LQ2,CMS-LQ3}, hadronic taus + jets + \MET~\cite{ATLAS-SUSY-taus} and multiple high-$p_T$ jets~\cite{ATLAS-6-7}, that would be more sensitive to these final states. We have also pointed out the value of designing a search based on the $t\bar t$ cross section measurement in the $\ell+\tau_h$ channel~\cite{CMS-ttbar-tau}.

At this moment in time, when theory does not provide a clear guidance as to how new physics will manifest itself, it is essential that all classes of final states are being covered experimentally.  Implementing the ideas presented here could greatly improve LHC reach, both for the RPV scenarios we discussed explicitly and for any new physics scenarios with similar final states.

\acknowledgments

This work was supported in part by DOE grant DE-FG02-96ER40959.

\end{document}